\documentclass[twoside]{dis07}
\usepackage[latin1]{inputenc}
\usepackage[dvips]{graphicx,epsfig,color}
\usepackage{wrapfig,rotating}
\usepackage{amssymb,amsmath,array}

\newcommand{\lsim}{\raisebox{-4pt}{%
    $\,\stackrel{\textstyle <}{\sim}\,$}}

\newcommand{\gev}{\operatorname{GeV}}

\newcommand{\ms}{\mskip 1.5mu}

\begin{document}
\title{Exclusive Meson Production at NLO}

\author{Markus Diehl and Wolfgang Kugler
%
\vspace{.3cm}\\
%
Deutsches Elektronen-Synchroton DESY \\
22603 Hamburg, Germany
}

\maketitle

\vspace{-13em}

\begin{flushright}
{\small DESY 07-093}
\end{flushright}

\vspace{10em}

\begin{abstract}
  We report on numerical studies of the NLO corrections to exclusive
  meson electroproduction, both in collider and fixed-target
  kinematics.  Corrections are found to be huge at small $x_B$ and
  sizeable at intermediate or large $x_B$.
\end{abstract}

\section{Motivation and general framework}

Generalized parton distributions (GPDs) are a versatile tool to
quantify important aspects of hadron structure in QCD.  They contain
unique information on the spatial distribution of partons
\cite{Burkardt:2002hr} and on the orbital angular momentum they carry
in the proton \cite{Ji:1996ek}.  The theoretically cleanest process
where GPDs can be studied is deeply virtual Compton scattering
(similar to inclusive DIS, which plays a dominant role in constraining
the usual parton densities).  Hard exclusive meson production is
harder to describe quantitatively, but it provides opportunities to
obtain important complementary constraints.  In particular, vector
meson production is more directly sensitive to the gluon
distributions, which enter the Compton amplitude only at
next-to-leading (NLO) order in $\alpha_s$.  Together with a wealth of
high-quality data \cite{Levy:2007}, this warrants efforts to bring
meson production under theoretical control as much as possible.

In the present contribution \cite{url} we investigate exclusive $\rho$
production ($\gamma^* p\to \rho\ms p$) using collinear factorization,
which is applicable in the limit of large photon virtuality $Q^2$ at
fixed Bjorken variable $x_B$ and fixed invariant momentum transfer $t$
to the proton \cite{Collins:1996fb}.  In practical terms, this means
that the description is restricted to sufficiently large $Q^2$ but can
be used for both small and large $x_B$, thus providing a common
framework for analyzing both collider and fixed-target data.  The
process amplitude can then be expressed in terms of GPDs for the
proton, the $q\bar{q}$ distribution amplitude for the $\rho$, and
hard-scattering kernels.  The kernels are known to NLO, i.e.\ to order
$\alpha_s^2$ \cite{Ivanov:2004zv}.

The requirement of ``sufficiently large'' $Q^2$ is demanding for meson
production.  Contributions that are formally suppressed by powers of
$1/Q^2$ cannot be calculated in a completely systematic way, but the
estimates
\cite{Goloskokov:2006hr,Vanderhaeghen:1999xj,Frankfurt:1995jw} agree
that for $Q^2$ of several $\gev^2$ the effect of the transverse quark
momentum inside the meson cannot be neglected in the hard-scattering
subprocess, as it is done in the collinear approximation.  This effect
can be incorporated in the modified hard-scattering picture
\cite{Goloskokov:2006hr,Vanderhaeghen:1999xj}, in color dipole models
\cite{Frankfurt:1995jw}, or in the MRT approach \cite{Martin:1996bp}.
Unfortunately, the calculation of $\alpha_s$ corrections remains not
only a technical but even a conceptual challenge in these approaches,
so that the perturbative stability of their results cannot be
investigated at present.  One strategy in this situation is to study
the NLO corrections in the collinear factorization framework,
identifying kinematical regions where they are moderate or small.
There one may use formulations incorporating power corrections from
transverse quark momentum with greater confidence.  This is the aim of
the present contribution.

In the following we show results for the convolution of the
unpolarized quark and gluon GPDs $H^q$ and $H^g$ with the
corresponding hard-scattering kernels and the asymptotic form of the
$\rho$ distribution amplitude.  We model the GPDs using a standard
ansatz based on double distributions \cite{Musatov:1999xp}, with the
CTEQ6M distributions as input.  Unless indicated explicitly, we take
$t=0$ and set the factorization and renormalization scales equal,
$\mu= \mu_F =\mu_R$.


\section{Numerical results}

\begin{figure}
\begin{center}
\includegraphics[width=0.49\columnwidth]{%
  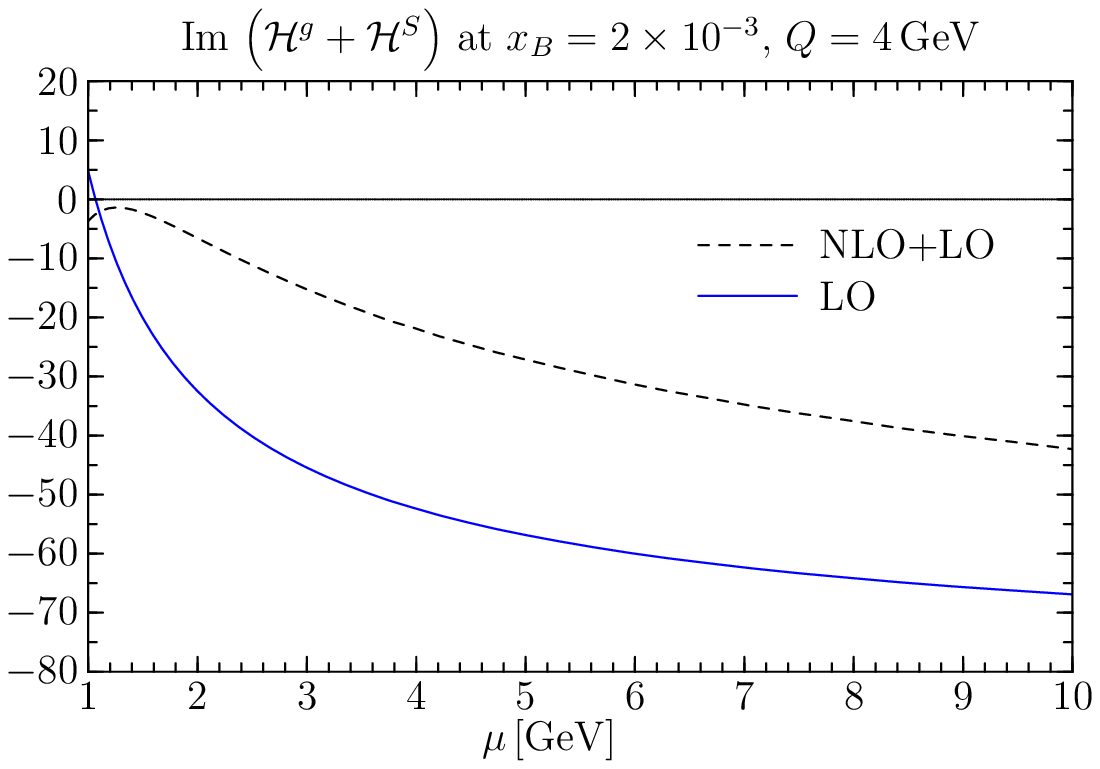}
\includegraphics[width=0.49\columnwidth]{%
  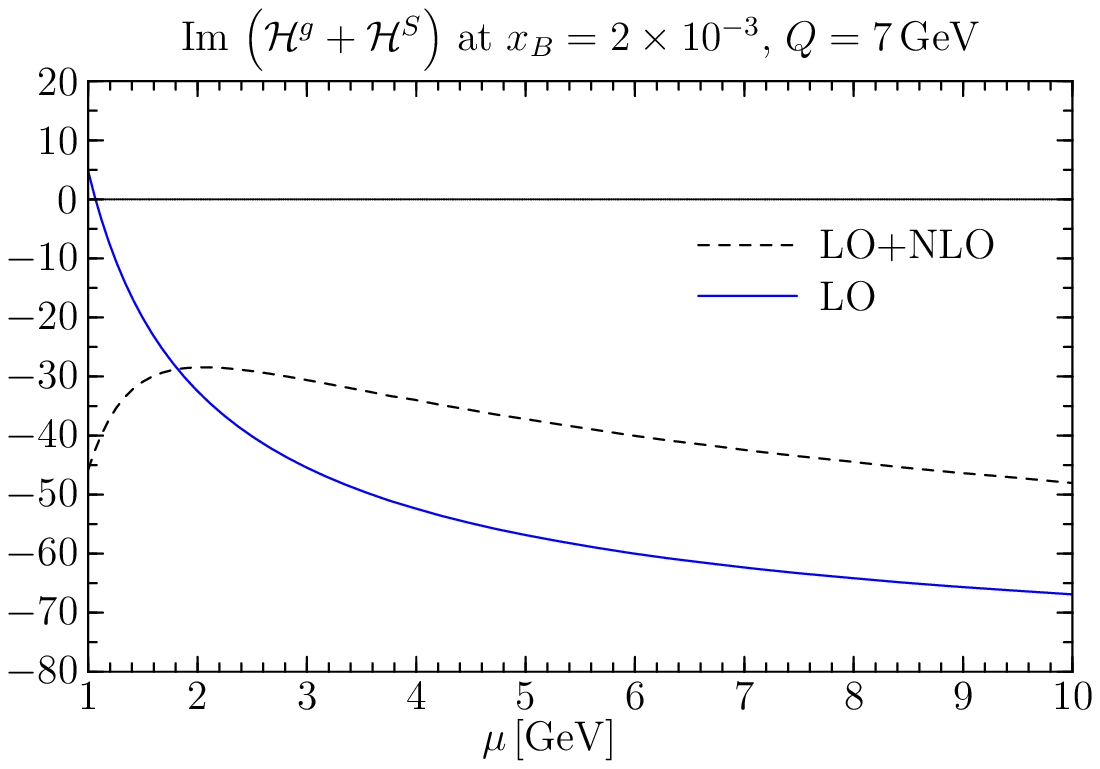}
\end{center}
\caption{\label{Fig:small-x} Imaginary part of the convolution
  integral for the sum of gluon and quark singlet distributions as a
  function of the renormalization and factorization scale $\mu$.}
\end{figure}

{

\begin{wrapfigure}{r}{0.5\columnwidth}
\centerline{\includegraphics[width=0.5\columnwidth]{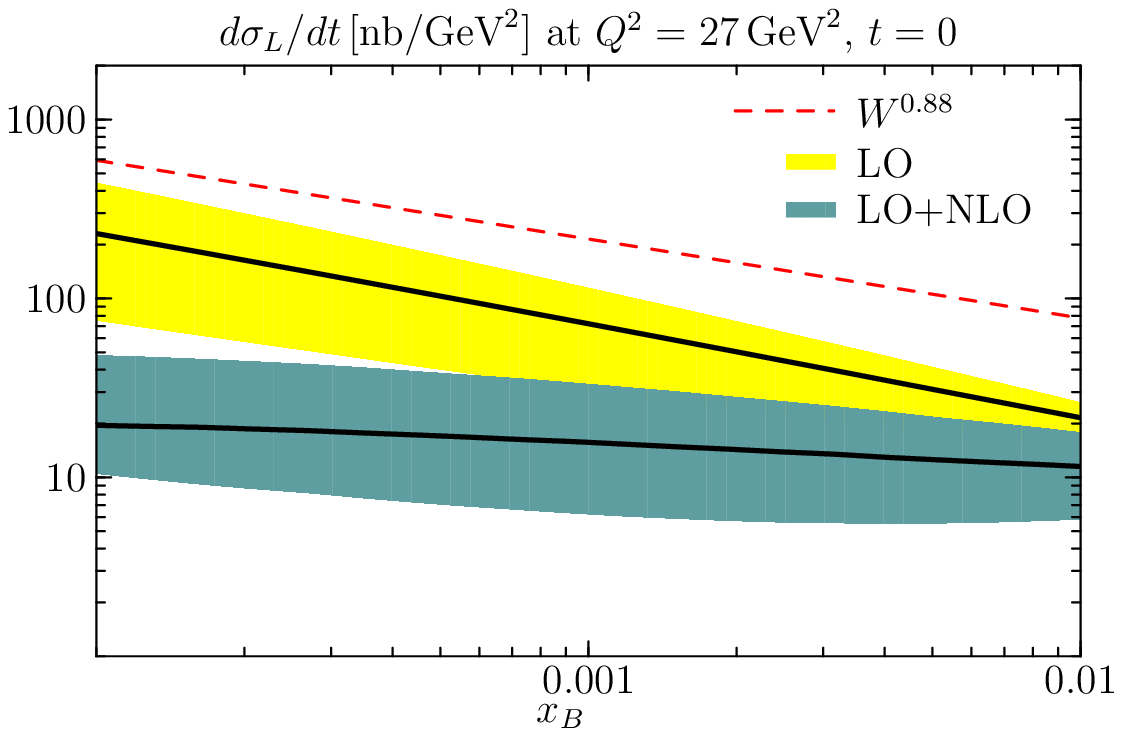}}
\caption{\label{Fig:zeus} Cross section for $\gamma^*p\to \rho\ms p$
  with a longitudinal photon.  Bands correspond to the range $Q/2 <
  \mu < 2Q$ and solid lines to $\mu=Q$.  We also show the power-law
  behavior $\sigma \propto W^{0.88}$ (with arbitrary normalization)
  obtained from a fit to data in the range $0.001 \protect\lsim x_B
  \protect\lsim 0.005$ \protect\cite{ZEUS:2001}.}
\end{wrapfigure}

In a wide kinematical range at small $x_B$, we find huge NLO
corrections which have opposite sign to the Born term and almost
cancel it.  This is shown for $x_B= 2\times 10^{-3}$ in
Fig.~\ref{Fig:small-x}, where there are indications for an onset of
perturbative stability at $Q= 7\gev$, but not yet at $Q= 4\gev$.
Taking $x_B= 2\times 10^{-4}$ one finds no stability even at $Q=
7\gev$, whereas for $x_B= 2\times 10^{-2}$ the corrections are of
tolerable size already at $Q= 4\gev$.

Figure~\ref{Fig:zeus} shows that in kinematics relevant for HERA
measurements, NLO corrections have a huge effect on the cross section
and moreover lead to a flat energy behavior in conflict with
experiment.  Due to the strong cancellations between LO and NLO terms,
the dependence on factorization and renormalization scale is not
reduced when going to NLO.

As already observed in \cite{Ivanov:2004zv} the large size of NLO
corrections at small $x_B$ can be traced back to BFKL-type logarithms
appearing first at NLO for vector meson production.  Such logarithms
are present in many processes (including DIS) but have a rather large
numerical prefactor in the present case.  It is to be hoped that
all-order resummation of these logarithms in the hard-scattering
kernel will give perturbative stability at small $x_B$.

}

\begin{figure}
\begin{center}
\includegraphics[width=0.49\columnwidth]{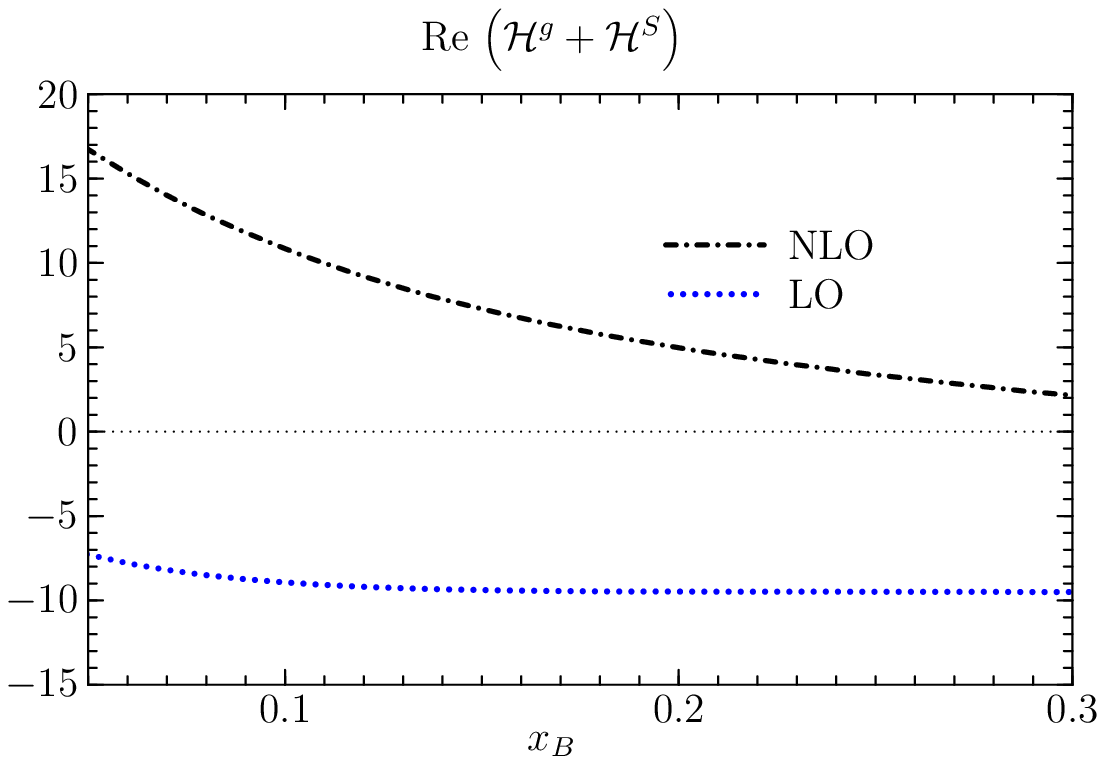}
\includegraphics[width=0.49\columnwidth]{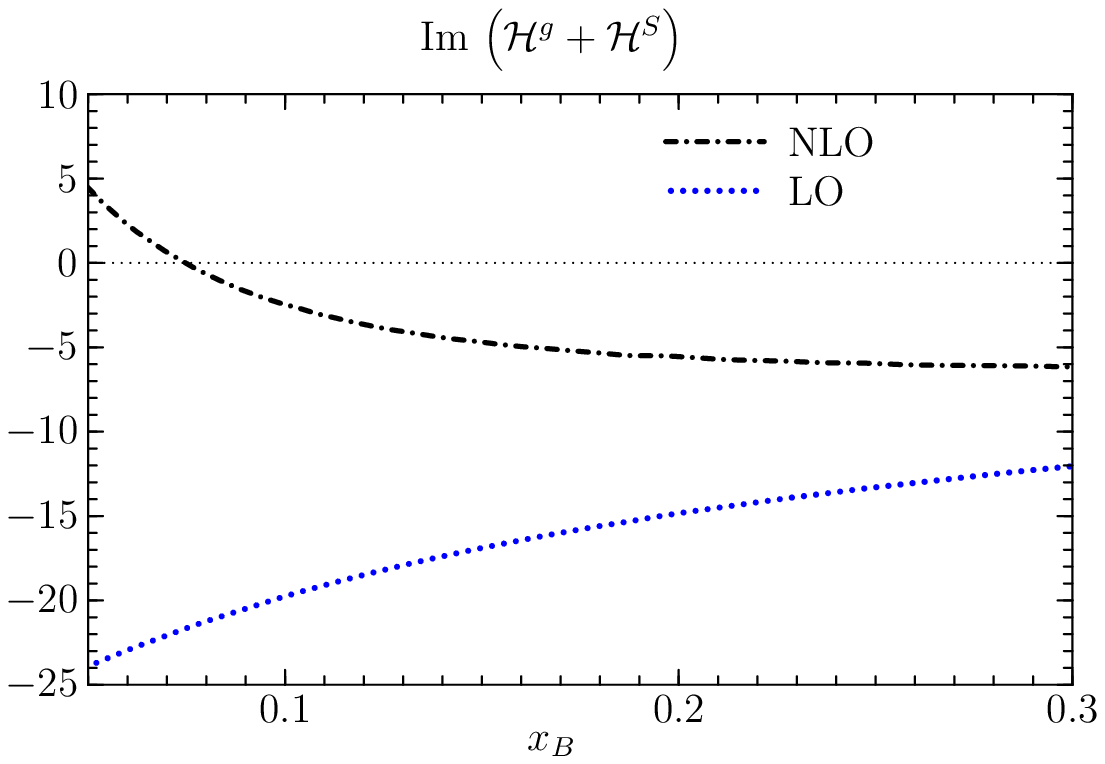}
\end{center}
\caption{\label{Fig:large-x} Real and imaginary part of the
  convolution integral for the sum of gluon and quark singlet
  distributions for $\mu_F=\mu_R=Q =2\gev$.}
\end{figure}

In the $x_B$ range relevant for experiments at COMPASS, HERMES, and
JLAB, we generally find corrections which are sizable but not huge.
An exception is the real part of the gluon and quark singlet
amplitudes, where corrections become large for decreasing $x_B$, as is
seen in the left panel of Fig.~\ref{Fig:large-x}.

\begin{figure}
\begin{center}
\includegraphics[width=0.49\columnwidth]{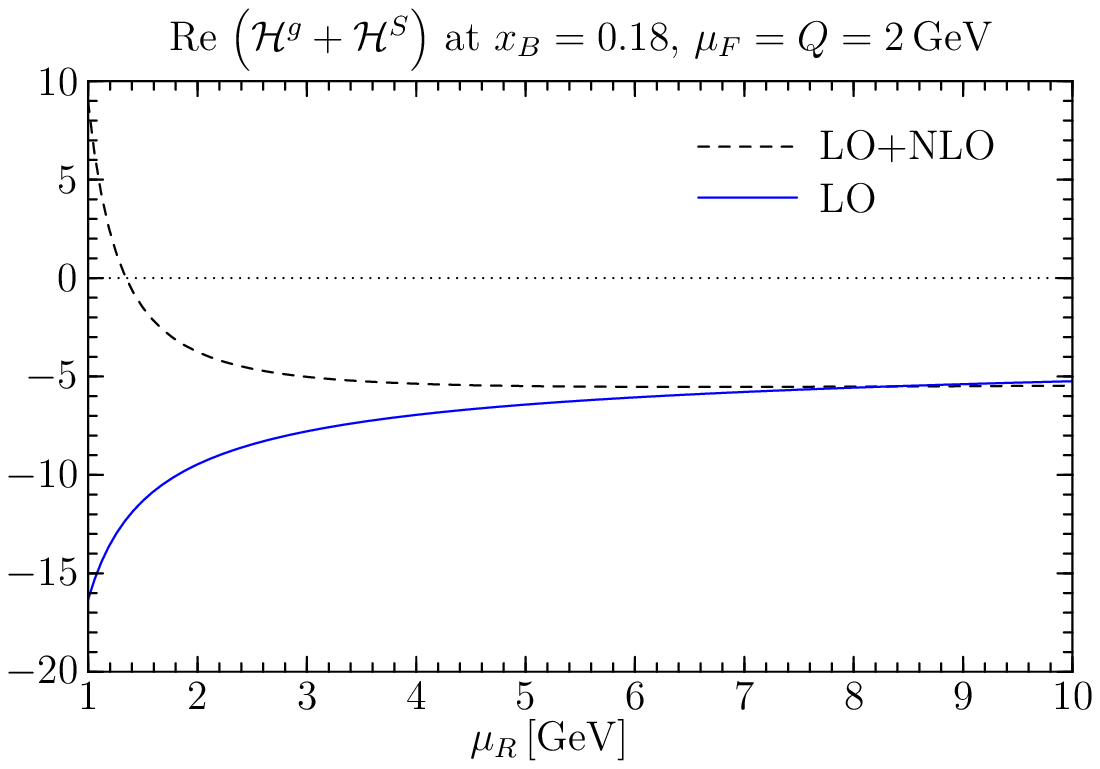}
\includegraphics[width=0.49\columnwidth]{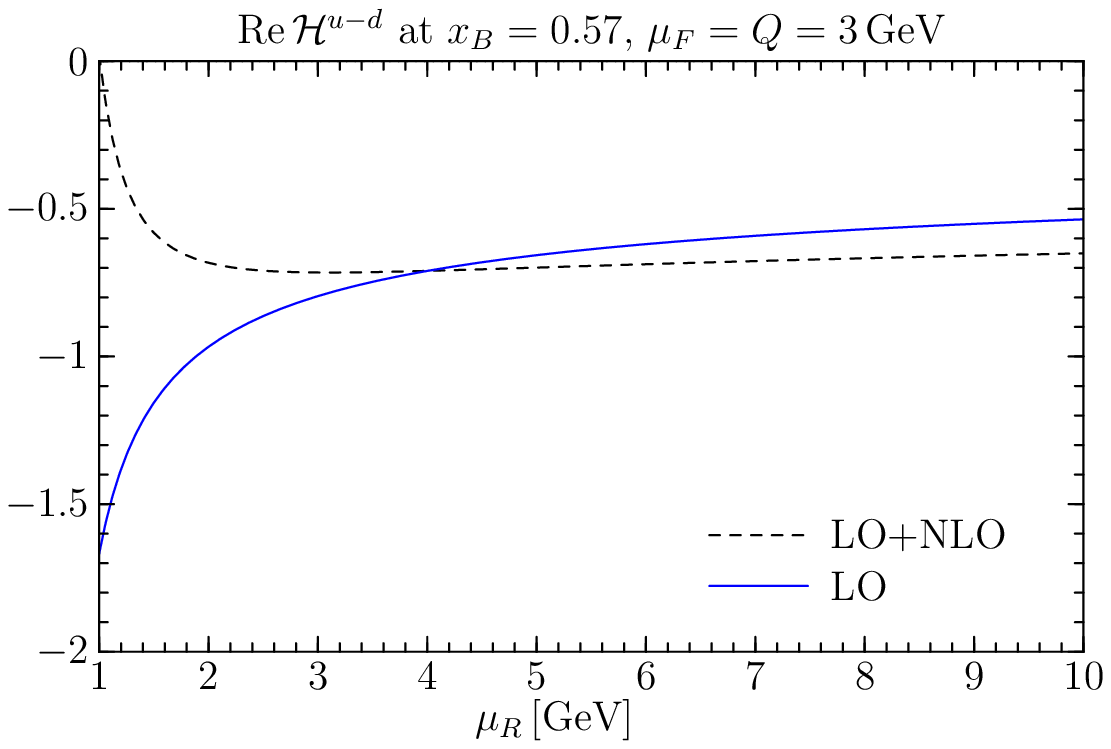}
\end{center}
\caption{\label{Fig:scale} Renormalization scale dependence of the
  real part of the convolution integrals for the sum of gluon and
  quark singlet distributions (left) and for for the difference of $u$
  and $d$ quark distributions (right).}
\end{figure}

In the quark nonsinglet sector there are large terms in the NLO kernel
due to gluon self-energy corrections.  The BLM procedure for setting
the renormalization scale aims at resumming these to all orders in
$\alpha_s$.  Applied to the process at hand, one finds however that
this requires $\mu_R$ to be typically an order of magnitude smaller
than $Q$ \cite{Belitsky:2001nq,Anikin:2004jb}.  This is outside the
validity of the perturbative calculation for most practically relevant
$Q$.  Numerically we find that for $\mu_R \lsim 2\gev$ the NLO
corrections become unstable for several convolution integrals, as
shown for examples in Fig.~\ref{Fig:scale}.

\begin{figure}
\begin{center}
\includegraphics[width=0.49\columnwidth]{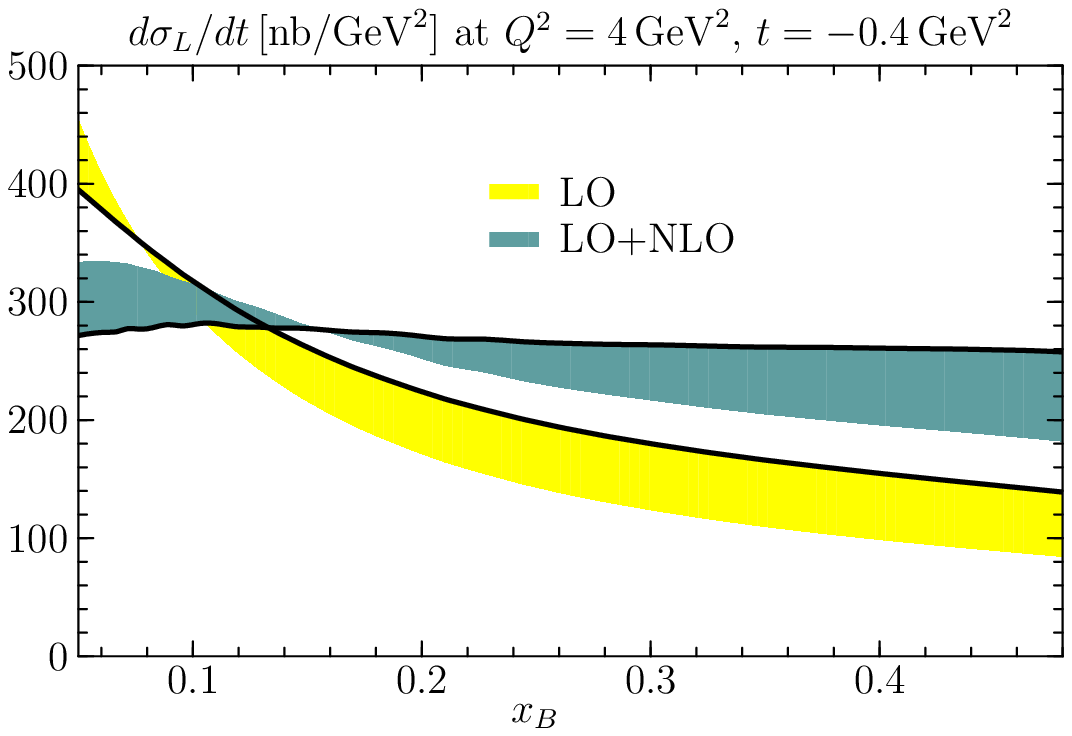}
\includegraphics[width=0.49\columnwidth]{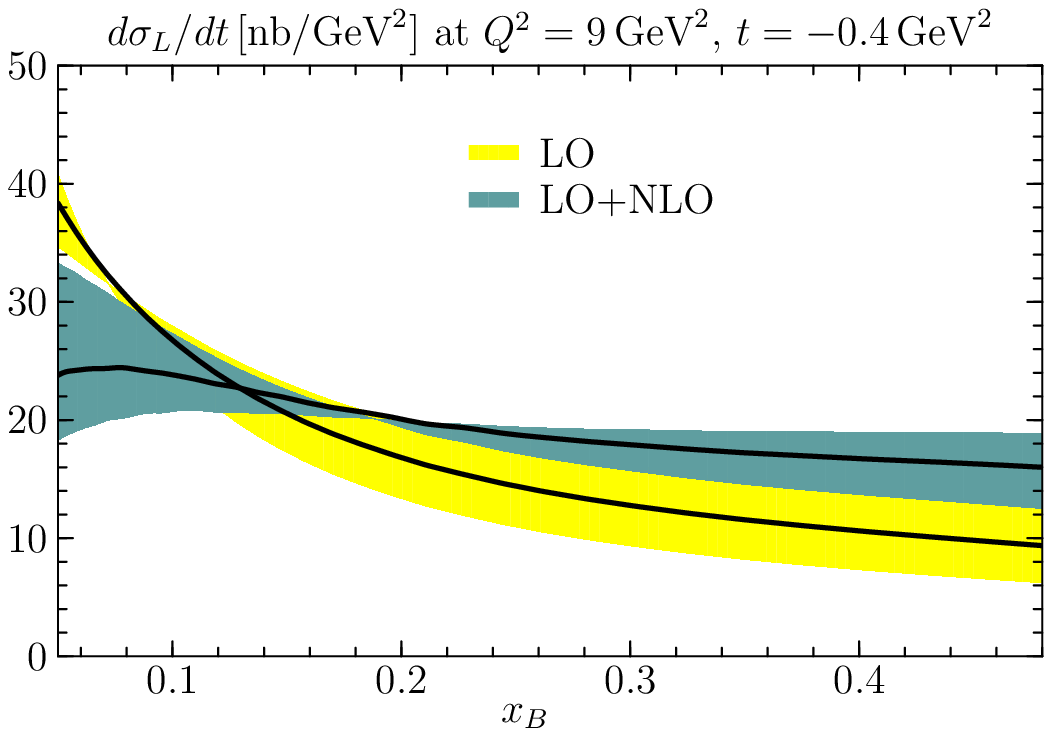}
\end{center}
\caption{\label{Fig:fixed-tar} Cross section for $\gamma^*p\to \rho\ms
  p$ with a longitudinal photon.  Bands correspond to the range $2\gev
  < \mu < 4\gev$ in the left and to $2\gev < \mu < 6\gev$ in the right
  plot, and solid lines to $\mu=Q$ in both cases. }
\end{figure}

We have therefore omitted this region when estimating the scale
setting error in Fig.~\ref{Fig:fixed-tar}, where we show the cross
section in typical fixed-target kinematics.  We see that NLO
corrections are quite large for $Q^2= 4\gev^2$, whereas for $Q^2=
9\gev^2$ and $x_B >0.1$ they become moderate.

\section*{Acknowledgments}

We gratefully acknowledge discussions with L.~Favart, H.~Fischer,
D.~Yu.~Ivanov, A.~Rostomyan and A.~Sch\"afer.  This work is supported
by the Helmholtz Association, contract number VH-NG-004.


\begin{footnotesize}

\end{footnotesize}


\end{document}